\documentclass[a4paper,11pt]{article}
\usepackage{pos}
\usepackage{natbib}

\title{Measurement of the energy-energy correlator in the back-to-back limit using the archived ALEPH $e^{+}e^{-}$ data at 91.2 GeV}

\author*[a]{Hannah Bossi}
\author[b]{Austin Baty}
\author[c]{Yi Chen}
\author[a]{Yu-Chen (Janice) Chen}
\author[a]{Gian-Michele Innocenti}
\author[d]{Marcello Maggi}
\author[a]{Chris McGinn}
\author[a]{Yen-Jie Lee}

\affiliation[a]{Massachusetts Institute of Technology, \\
  Cambridge, MA, 02139}

\affiliation[b]{University of Illinois Chicago,\\
1200 West Harrison St. Chicago, Illinois 60607}

\affiliation[c]{Vanderbilt University, \\Nashville, TN 37235}

\affiliation[d]{
INFN Sezione di Bari, \\Bari, Italy}

\emailAdd{hannah.bossi@cern.ch}

\abstract{Recently, energy-energy correlators (EECs) have garnered renewed interest for studying hadronic collisions at the Large Hadron Collider (LHC) and the Relativistic Heavy Ion Collider (RHIC). EEC measurements within jets provide a clear scale separation, facilitating the study of both perturbative and non-perturbative Quantum Chromodynamics (QCD) in the collinear limit. These proceedings present recent EEC results from the archived ALEPH $e^{+}e^{-}$ data taken at LEP at $\sqrt{s}$ = 91.2 GeV. In $e^{+}e^{-}$ collisions, perturbative and non-perturbative QCD can be studied with EECs in both the collinear limit using jets and the back-to-back limit using all particles as well as the transition between these two regimes. Comparisons of these results to generators and future extensions of this work will also be discussed. 
}

\FullConference{%
  12th Edition of the Large Hadron Collider Physics Conference,\\
  3-7 June, 2024 \\
  Boston, Massachusetts, USA
}

\begin{document}
\maketitle

\section{Introduction}
The $e^{+}e^{-}$, collision environment provides the cleanest setting for studying QCD, as the colliding objects are fundamental particles. Unlike hadronic collisions, $e^{+}e^{-}$ collisions avoid complications such as beam remnants, gluonic initial-state radiation, and parton distribution functions. In recent years, re-analysis of the archived data from the Apparatus for LEp PHysics (ALEPH) experiment at the Large Electron Positron Collider (LEP) has offered a pathway to investigate current open questions in a well-understood system. For example, two particle correlations reanalyzed with LEP 1~\cite{Badea:2019vey} and LEP 2~\cite{Chen:2023njr} data reveal an excess in the highest multiplicity interval of LEP 2 data not seen in simulation, providing new insights into the presence of flow signatures in small systems. The archived data enables the application of modern experimental techniques developed in recent years. One recent example of this uses jet reconstruction algorithms developed in the early 2000s~\cite{Cacciari:2008gp} to looks at jets and their substructure in ALEPH data~\cite{Chen:2021uws} serving as a precision tests of analytical calculations and phenomenological models in the collinear limit of QCD.

In recent years the N-point energy correlation functions, a class of observables originally developed for the study of $e^{+}e^{-}$ collisions~\cite{PhysRevLett.41.1585,Basham:1978zq, Basham:1979gh,Basham:1977iq}, has been re-imagined in hadronic collision environments~\cite{CMS:2024mlf,ALICE:2024dfl,Tamis:2023guc}. The N-point energy correlation functions are defined as the correlation of the energy flow operator
\begin{equation}
    \langle \Psi|\mathcal{E}(\vec n_1) \mathcal{E}(\vec n_2)\cdots \mathcal{E}(\vec n_k)|\Psi \rangle
\end{equation}
where $ \mathcal{E}(\vec n_1) = 
    \lim_{r\rightarrow \infty} \int \mathrm{d}t r^2 n_1^i T_{0i}(t,r\vec{n}_1)$.
ENCs characterize the energy flow in the direction of the vector $\hat{n}$, which is commonly replaced by a distance variable such as $\Delta R = \sqrt{\Delta y^{2} + \Delta\phi^{2}}$ or $\theta_{\rm L}$ (the opening angle in radians) in collider environments. In experimental contexts it is also common to measure the \textit{projected} correlators that integrate out all shape information keeping the longest side (herein referred to as $R_{\rm L}$ or $\theta_{\rm L}$) fixed. The projected correlators are useful to isolate the scaling behavior, offering a clean separation between the free hadron region, the quark/gluon region, and the transition between these two regimes. The full energy correlators are useful for studying the \textit{shape} of the energy flow, which has recently been studied in simulated heavy-ion collisions~\cite{Andres:2022ovj,Andres:2023xwr,Andres:2024ksi,Barata:2023bhh,Yang:2023dwc,Bossi:2024qho,Andres:2024hdd}. 

This paper presents the first fully corrected measurement of the projected two-point energy correlators (E2C). The clean environment of $e^{+}e^{-}$ collisions enables the use of all particles in the event, not only those within a jet, allowing the measurement to probe the collinear limit of QCD where $\theta_{\rm L} \sim 0$ all the way to the back-to-back (or Sudakov) limit where $\theta_{\rm L}\sim \pi$. Here the E2C is measured as a function of $z$, where $z = \frac{1-\text{cos}(\theta_{\rm L})}{2}$, 
in order to compare the free hadron region in the collinear and back to back limits. In this definition the E2C can be written as 
\begin{equation}
    \text{E2C}(z) = \sum^{n}_{\rm i,j} d\sigma \frac{E_{\rm i}E_{\rm j}}{E^{2}} \delta(z - z_{\rm ij}).
\end{equation}
In order to perform a fully-corrected measurement, the E2C must be corrected for detector effects and energy smearing in $z$ and $E_{\rm i}E_{\rm j}$ using a two-dimensional Bayesian unfolding procedure~\cite{D'Agostini:265717}. Here, the energy scale $E$ is fixed at the collision energy of $\sqrt{s} = 91.2 \text{ GeV}$, eliminating the need to explicitly correct for the energy scale and significantly simplifying the unfolding procedure. The remainder of these proceedings is outlined as follows. The details of the dataset utilized for this measurement will be discussed in Section~\ref{sec:dataset}. The results will be presented in Section~\ref{sec:results}. Finally, some conclusions and future research directions will be presented in Section~\ref{sec:conclusions}.

\section{ALEPH Archived Dataset}~\label{sec:dataset}
This measurement uses data collected during the LEP 1 period with ALEPH~\footnote{Currently only data taken during 1994 is analyzed due to the availability of simulation.}, later archived in the MIT open data format~\cite{Tripathee:2017ybi}~\footnote{The authors would like to specifically thank Roberto Tenchini and Guenther Dissertori for their help in making this archival effort possible through their advice and expertise with ALEPH data.}. ALEPH was located at IP4 of the LEP ring and features a complex detector design useful for particle identification and a measurements of particle properties. This was done using precision tracking via its time projection chamber (TPC) and inner tracking chamber and calorimetry via its hadronic and electromagnetic calorimeters. The tracking detectors of ALEPH are located inside of a superconducting magnet with a field strength of 1.5 T.  The LEP 1  dataset was taken at the Z-pole with a center of mass energy of  $\sqrt{s} = 91.2 \text{ GeV}$, where the dominant process in hadronic data is $e^{+}e^{-} \rightarrow q\bar{q}$. The event selection for this analysis requires at least five high-quality tracks and a total reconstructed charged-particle energy of at least 15 GeV. In addition events must also have a sphericity axis with a polar angle that falls into the acceptance window of $7\pi/36 \leq \theta_{\rm sphericity} \leq 29\pi/36$. For the E2C measurement, only charged particles were included, with tracks required to have at least 4 TPC hits and a transverse momentum above 0.2 GeV. These tracks must also fall in an acceptance of $|\text{cos}(\theta)| < 0.94$ and have an impact parameter within a radial distance of  $d_{\rm 0} < 2 \text{ cm}$ and a longitudinal distance  $z_{\rm 0} < 10 \text{ cm}$ from the interaction point.

\section{Results}~\label{sec:results}
The results are presented using a unique double-log format to emphasize angular limits, with the collinear region spanning $z\sim0$ to $z\sim1/2$ and the Sudakov region spanning $z \sim 1$ to $z \sim 1 - 1/2$. The E2C in the collinear limit is similar to the analogous quantity constructed from particles inside jets with clear free hadron, transition, and quark/gluon regions. However, in this case there is no cutoff in the distribution due to the jet radius. Then the Sudakov region demonstrates a reflected version of these regions with the free hadron region appearing at large values of $z$ and then the transition and quark/gluon region emerge as $z$ decreases. 

This initial result iteration provides conservative systematic uncertainty estimates, with expectations for further refinement. The dominant systematic uncertainties  are those associated with the unfolding procedure, including the choice of binning, the number of iterations, as well as the impact of the choice of prior. An uncertainty for the TPC hit requirement for good tracks (taking a nominal value of 4 and increased to 7 for the systematic uncertainty) as well as for the matching method are also included.   

\begin{figure}[ht!]
    \centering
    \includegraphics[width=0.49\linewidth]{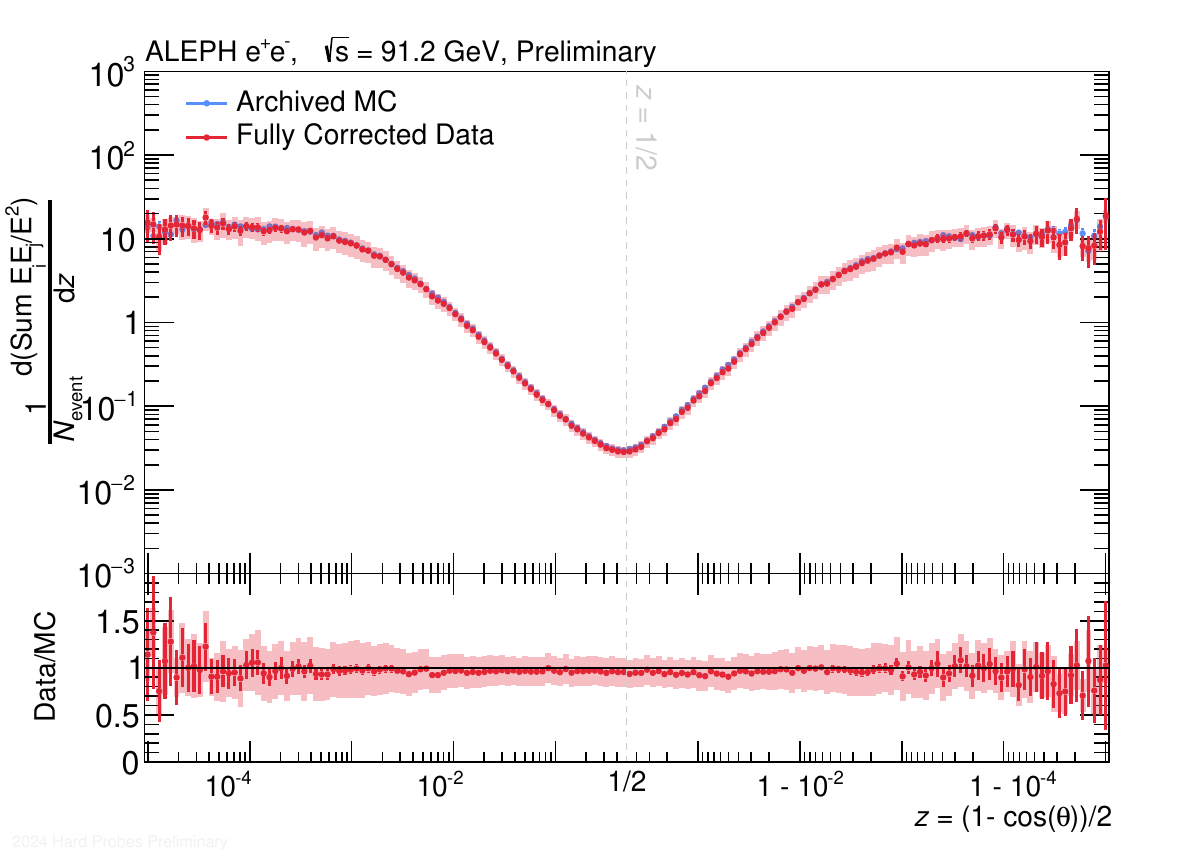}
    \includegraphics[width=0.49\linewidth]{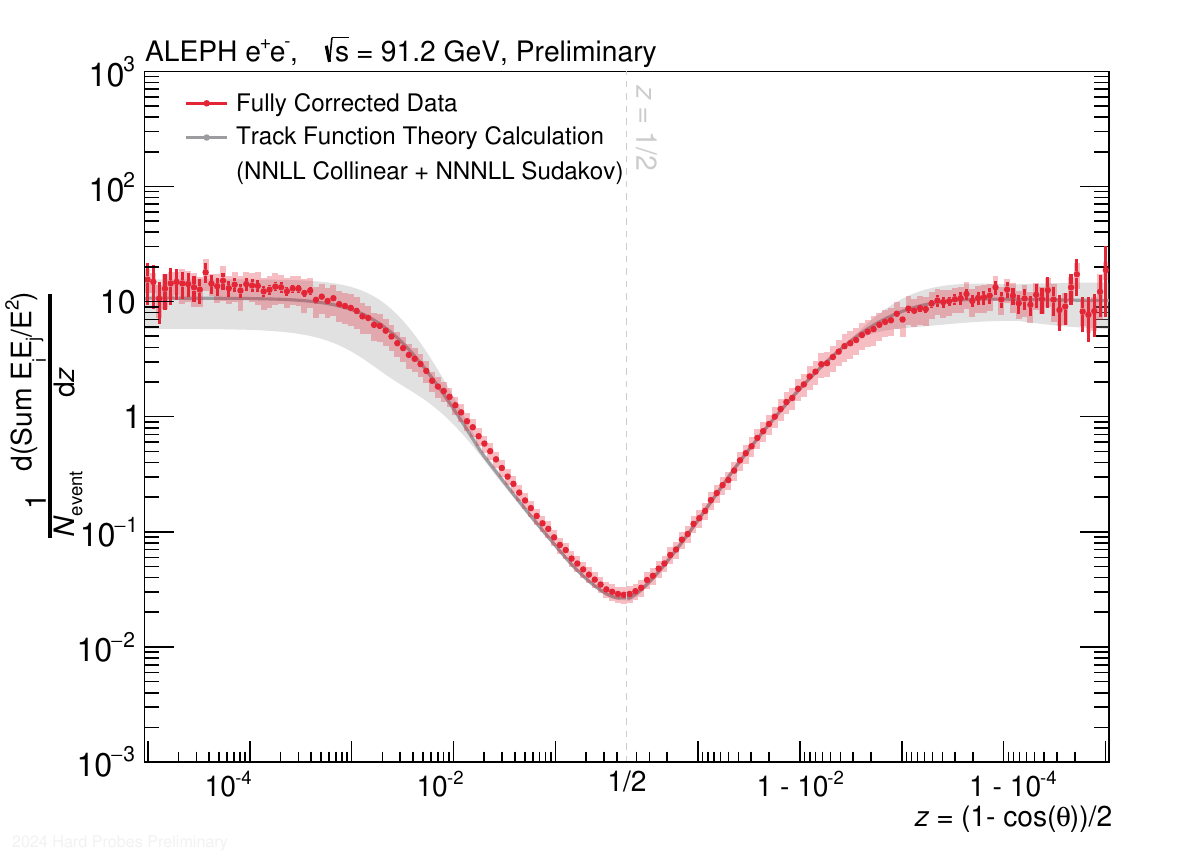}
    \caption{Left: E2C distribution as a function of $z$ for the archived PYTHIA 6 MC distribution (blue) and the fully corrected ALEPH data with the corresponding systematic uncertainties (red). The ratio of the data to MC is shown in the bottom panel. Right: E2C distributions as a function of $z$ for the fully corrected data compared to a track function theory calculation with NNLL Collinear and NNNLL Sudakov regions.}
    \label{fig:results}
\end{figure}

In the left panel of Figure \ref{fig:results} the fully-corrected E2C is shown as a function of $z$ compared to the archived PYTHIA 6 MC, with the ratio of the two distributions shown in the bottom panel. The data demonstrates excellent agreement with the MC simulation, although significant differences exist between generators for jet substructure observables~\cite{Chen:2021uws}. Additionally, with the unprecedented binning and kinematic reach, one can see that the collinear and Sudakov free hadron regions are roughly compatible with one another.~\footnote{Quantitative comparisons between the two are left for future work.} In the right panel of Figure \ref{fig:results}, this distribution is compared to a theoretical calculation~\footnote{"A Precision Calculation of the Energy-Energy Correlator on Tracks", Max Jaarsma, Yibei Li, Ian Moult, Wouter Waalewijn, HuaXing Zhu, in preparation}. In the collinear region this calculation is performed as a Next-to-Next-To-Leading-Log (NNLL) collinear resummation. In the Sudakov region this is implemented as a Next-to-Next-To-Next-To-Leading-Log (NNLL) Sudakov resummation where the Collins-Soper Kernel is extracted from lattice QCD. In both predictions the non-perturbative parameter $\Omega$ is extracted from the thrust distribution. The theoretical calculation exhibits excellent agreement with the data over all regions of phase space. This measurement represents the first of its kind and will be useful to further constrain the theory in the relatively unexplored Sudakov limit.

\section{Conclusions}~\label{sec:conclusions}
These proceedings present the first fully-corrected measurement spanning from the collinear to the back-to-back limit of QCD using ALEPH archived data. These studies show excellent agreement with the archived PYTHIA 6 MC and theoretical calculations, providing crucial tests for QCD calculations and phenomenological models. This is especially true in the relatively-unexplored Sudakov limit where this measurement provides one of the first experimental constraints. 

The ALEPH archived dataset is full of nearly-limitless opportunities. For example, a value for $\alpha_{\rm s}$ can be extracted from the ratio of higher point correlators to the E2C. The $\alpha_{\rm s}$ fits from $e^{+}e^{-}$ event shapes and analytical hadronization were recently removed from the world average~\cite{Huston:2023ofk}, making such a measurement timely. This measurement marks the beginning of a new investigative direction in $e^{+}e^{-}$ collisions, revisiting concepts from the 1970s to address contemporary physics questions. This work also has the potential to shape the future, serving as a catalyst to inspire and inform studies at the proposed FCC-ee~\cite{Benedikt:2651299}.

\bibliographystyle{unsrt}
\bibliography{EEC}

\begin{thebibliography}{10}

\bibitem{Badea:2019vey}
Anthony Badea, Austin Baty, Paoti Chang, Gian~Michele Innocenti, Marcello Maggi, Christopher Mcginn, Michael Peters, Tzu-An Sheng, Jesse Thaler, and Yen-Jie Lee.
\newblock {Measurements of two-particle correlations in $e^+e^-$ collisions at 91 GeV with ALEPH archived data}.
\newblock {\em Phys. Rev. Lett.}, 123(21):212002, 2019.

\bibitem{Chen:2023njr}
Yu-Chen Chen et~al.
\newblock {Long-range near-side correlation in $e^+e^-$ collisions at 183-209 GeV with ALEPH archived data}.
\newblock {\em Phys. Lett. B}, 856:138957, 2024.

\bibitem{Cacciari:2008gp}
Matteo Cacciari, Gavin~P. Salam, and Gregory Soyez.
\newblock {The anti-$k_t$ jet clustering algorithm}.
\newblock {\em JHEP}, 04:063, 2008.

\bibitem{Chen:2021uws}
Yi~Chen et~al.
\newblock {Jet energy spectrum and substructure in $e^+e^-$ collisions at 91.2 GeV with ALEPH Archived Data}.
\newblock {\em JHEP}, 06:008, 2022.

\bibitem{PhysRevLett.41.1585}
C.~Louis Basham, Lowell~S. Brown, Stephen~D. Ellis, and Sherwin~T. Love.
\newblock Energy correlations in electron-positron annihilation: Testing quantum chromodynamics.
\newblock {\em Phys. Rev. Lett.}, 41:1585--1588, Dec 1978.

\bibitem{Basham:1978zq}
C.L. Basham, L.S. Brown, S.D. Ellis, and S.T. Love.
\newblock {Energy Correlations in electron-Positron Annihilation in Quantum Chromodynamics: Asymptotically Free Perturbation Theory}.
\newblock {\em Phys. Rev. D}, 19:2018, 1979.

\bibitem{Basham:1979gh}
C.~Louis Basham, Lowell~S. Brown, Stephen~D. Ellis, and Sherwin~T. Love.
\newblock {Energy Correlations in Perturbative Quantum Chromodynamics: A Conjecture for All Orders}.
\newblock {\em Phys. Lett. B}, 85:297--299, 1979.

\bibitem{Basham:1977iq}
C.~Louis Basham, Lowell~S. Brown, S.~D. Ellis, and S.~T. Love.
\newblock {Electron - Positron Annihilation Energy Pattern in Quantum Chromodynamics: Asymptotically Free Perturbation Theory}.
\newblock {\em Phys. Rev. D}, 17:2298, 1978.

\bibitem{CMS:2024mlf}
Aram Hayrapetyan et~al.
\newblock {Measurement of energy correlators inside jets and determination of the strong coupling $\alpha_\mathrm{S}(m_\mathrm{Z})$}.
\newblock 2 2024.

\bibitem{ALICE:2024dfl}
Shreyasi Acharya et~al.
\newblock {Exposing the parton-hadron transition within jets with energy-energy correlators in pp collisions at $\sqrt{\textit s}=5.02$ TeV}.
\newblock 9 2024.

\bibitem{Tamis:2023guc}
Andrew Tamis.
\newblock {Measurement of Two-Point Energy Correlators Within Jets in $pp$ Collisions at $\sqrt{s}$ = 200 GeV at STAR}.
\newblock In {\em {11th International Conference on Hard and Electromagnetic Probes of High-Energy Nuclear Collisions}: {Hard Probes 2023}}, 9 2023.

\bibitem{Andres:2022ovj}
Carlota Andres, Fabio Dominguez, Raghav Kunnawalkam~Elayavalli, Jack Holguin, Cyrille Marquet, and Ian Moult.
\newblock {Resolving the Scales of the Quark-Gluon Plasma with Energy Correlators}.
\newblock {\em Phys. Rev. Lett.}, 130(26):262301, 2023.

\bibitem{Andres:2023xwr}
Carlota Andres, Fabio Dominguez, Jack Holguin, Cyrille Marquet, and Ian Moult.
\newblock {A coherent view of the quark-gluon plasma from energy correlators}.
\newblock {\em JHEP}, 09:088, 2023.

\bibitem{Andres:2024ksi}
Carlota Andres, Fabio Dominguez, Jack Holguin, Cyrille Marquet, and Ian Moult.
\newblock {Towards an Interpretation of the First Measurements of Energy Correlators in the Quark-Gluon Plasma}.
\newblock 7 2024.

\bibitem{Barata:2023bhh}
Jo\~ao Barata, Paul Caucal, Alba Soto-Ontoso, and Robert Szafron.
\newblock {Advancing the understanding of energy-energy correlators in heavy-ion collisions}.
\newblock {\em JHEP}, 11:060, 2024.

\bibitem{Yang:2023dwc}
Zhong Yang, Yayun He, Ian Moult, and Xin-Nian Wang.
\newblock {Probing the Short-Distance Structure of the Quark-Gluon Plasma with Energy Correlators}.
\newblock {\em Phys. Rev. Lett.}, 132(1):011901, 2024.

\bibitem{Bossi:2024qho}
Hannah Bossi, Arjun~Srinivasan Kudinoor, Ian Moult, Daniel Pablos, Ananya Rai, and Krishna Rajagopal.
\newblock {Imaging the Wakes of Jets with Energy-Energy-Energy Correlators}.
\newblock 7 2024.

\bibitem{Andres:2024hdd}
Carlota Andres, Jack Holguin, Raghav Kunnawalkam~Elayavalli, and Jussi Viinikainen.
\newblock {Minimizing Selection Bias in Inclusive Jets in Heavy-Ion Collisions with Energy Correlators}.
\newblock 9 2024.

\bibitem{D'Agostini:265717}
Giulio D'Agostini.
\newblock {A multidimensional unfolding method based on Bayes' Theorem}.
\newblock Technical report, DESY, Hamburg, 1994.

\bibitem{Tripathee:2017ybi}
Aashish Tripathee, Wei Xue, Andrew Larkoski, Simone Marzani, and Jesse Thaler.
\newblock {Jet Substructure Studies with CMS Open Data}.
\newblock {\em Phys. Rev.}, D96(7):074003, 2017.

\bibitem{Huston:2023ofk}
Joey Huston, Klaus Rabbertz, and Giulia Zanderighi.
\newblock {Quantum Chromodynamics}.
\newblock 12 2023.

\bibitem{Benedikt:2651299}
Michael Benedikt, Alain Blondel, Olivier Brunner, Mar Capeans~Garrido, Francesco Cerutti, Johannes Gutleber, Patrick Janot, Jose~Miguel Jimenez, Volker Mertens, Attilio Milanese, Katsunobu Oide, John~Andrew Osborne, Thomas Otto, Yannis Papaphilippou, John Poole, Laurent~Jean Tavian, and Frank Zimmermann.
\newblock {FCC-ee: The Lepton Collider: Future Circular Collider Conceptual Design Report Volume 2. Future Circular Collider}.
\newblock Technical Report~2, CERN, Geneva, 2019.

\end{thebibliography}
\end{document}